\documentclass[english]{llncs}
\usepackage[]{fontenc}
\usepackage[latin1]{inputenc}
\usepackage{babel}
\usepackage{times}
\usepackage{graphicx}
\usepackage{subfigure}
\begin{document}
\title{3-D Ultrasound Probe Calibration\\for Computer-Guided Diagnosis and Therapy}
\author{Michael Baumann\inst{1,2} \and Vincent Daanen\inst{1} \and Antoine Leroy\inst{2} \and Jocelyne Troccaz\inst{1}}
\institute{TIMC Laboratory, GMCAO Department, Institut d'Ingénierie de l'Information de Santé (IN3S), Faculty of Medecine - F-38706 La Tronche cedex, France, \email{michael.baumann@imag.fr}
\and KOELIS, 4, Avenue de l'Obiou - F-38700 La Tronche cedex, France}
\authorrunning{Michael Baumann et al.}
\tocauthor{Michael Baumann (TIMC laboratory, KOELIS),
Vincent Daanen (TIMC laboratory),
Antoine Leroy (KOELIS),
Jocelyne Troccaz (TIMC laboratory)}
\maketitle

\begin{abstract}
With the emergence of swept-volume ultrasound (US) probes, precise and almost real-time US volume imaging has become available. This offers many new opportunities for computer guided diagnosis and therapy, 3-D images containing significantly more information than 2-D slices. However, computer guidance often requires knowledge about the exact position of US voxels relative to a tracking reference, which can only be achieved through probe calibration. In this paper we present a 3-D US probe calibration system based on a membrane phantom. The calibration matrix is retrieved by detection of a membrane plane in a dozen of US acquisitions of the phantom. Plane detection is robustly performed with the 2-D Hough transformation. The feature extraction process is fully automated, calibration requires about 20 minutes and the calibration system can be used in a clinical context. The precision of the system was evaluated to a root mean square (RMS) distance error of 1.15mm and to an RMS angular error of 0.61°. The point reconstruction accuracy was evaluated to 0.9mm and the angular reconstruction accuracy to 1.79°.
\end{abstract}

\section{Introduction}
Until recently, 3-D ultrasound (US) volumes had to be manually reconstructed from a number of 2-D US slices acquired while slowly moving a 2-D probe over the target region. The so-called \textit{3D freehand method} is time-consuming, imprecise and not usable for many clinical applications requiring real-time acquisition \cite{Rohling99Comparison}. The emergence of 3-D swept-volume US probes solved most of the enumerated problems: a mechanical device capable of sweeping the 2-D crystal array of the probe over a target region makes it possible to acquire 3-D US volumes accurately and almost in real-time (1s to 4s per acquisition)\footnote{However, most currently available systems don't yet provide a real-time data transfer interface for 3-D data. Nevertheless one can acquire so called "4-D" images of three orthogonal volume slices in real-time using a video-capture device. In the rest of this article we make abstraction of this restriction, hoping that it will disappear with the next generation of 3-D echographs.}.

These new capabilities open an entire new field of applications in the domain of computer guided medical interventions based on US imaging. One can imagine tool guidance systems that would operate with permanently updated US volumes, visualizing for instance slices at the tool tip position. More sophisticated applications could carry out target localization inside the volumes through real-time registration and segmentation techniques, thus allowing to match pre-operative planning with intra-operative data.

However, US-based guidance often requires knowledge about the position and orientation of the US volume in space. When using a tracking system this can be achieved by calibrating the US acquisition volume with a tracking reference fixed on the probe. Unfortunately it is virtually impossible to derive the calibration parameters directly from the geometry and parameterization of the probe. Almost all existing calibration systems rely therefore on statistical or segmentation-based object matching methods.

\subsection{Calibration Methods Overview}
A variety of techniques for 2-D US calibration was proposed in the literature; a comprehensive review being given in \cite{Mercier04CalibReview}. Calibration methods can be classified with respect to the target geometry they rely on. \textit{Single-point target} methods identify a point, i.e. a bead, a calibrated pointer tip or a cross-wire, in the US image \cite{State94CaseStudy,Prager98RapidCalibFreehand,Muratore01BeamCalib}. The difficulties consist in automatic geometry extraction in the US slice and US beam alignment with the phantom. 

\textit{Multi-point target} phantoms are extensions of the single-point bead or cross-wire phantoms. They consist of a number of point targets with precisely known coordinates in phantom space. Their geometric configuration makes it possible to derive the calibration parameters from the distances between the reconstructed intersection points visible in the 2-D US scan \cite{Leotta04Efficient}. Compared to single-point phantoms they require less image acquisitions due to their more discriminative geometry, but share the phantom alignment and feature extraction problems. 

\textit{Z-fiducial} or \textit{N-fiducial} phantoms address the alignment problem of point target methods. A calibration point is determined from the intersection points of a number of nylon strings with the US beam. This is possible due to a sufficiently discriminative wire geometry \cite{Comeau98NeuroSurg,Pagoulatos01Fast,Lindseth03Probe}. Fiducial methods are more robust than point target methods but the difficulties concerning fully automatic feature extraction subsist. Also, Z- or N-fiducial phantoms require a high manufacturing accuracy to achieve a satisfying calibration quality.

\textit{Wall phantom} methods are based on detection of the intersecting line of a planar surface with the 2-D US beam. In \cite{Prager98RapidCalibFreehand}, a water tank bottom is imaged for calibration. The authors of \cite{Lango00USGuidedSurgery} address the reverberation and line thickness problems inherent of wall phantoms by using a membrane variant. Both phantoms have difficulties when confronted with steep angles between the US beam and the plane because they cause line intensity and line sharpness degradation \cite{Mercier04CalibReview}. The Cambridge phantom scans a rotating bar, thus creating a virtual plane, to solve these problems \cite{Prager98RapidCalibFreehand}. The advantage of plane phantoms lies in the robustness of the feature extraction process which can, as a consequence, be reliably and fully automated. The pitfall of this method lies in the non-discriminative phantom geometry which can result in underdetermined systems if the acquired calibration samples do not cover all degrees of freedom. This can be avoided by strictly respecting the acquisition protocols presented in \cite{Prager98RapidCalibFreehand,Treece03HighDefinition}. 

\textit{Registration Phantoms}: the last class of calibration methods relies on surface or intensity based registration techniques and therefore has the advantage of being independent of phantom geometry. The only requirement on phantom shape is that its US image is sufficiently discriminative with respect to rotations and translations, which is true for non-symmetric phantoms. The lack of precision of registration algorithms is the major drawback of this approach. To our knowledge, only one study examined registration-based 2-D probe calibration, registering US slices with an MRI image of the phantom \cite{Blackall00Registration}. A 3-D approach is discussed in the next paragraph.

\subsection{3-D Probe Calibration}
Until today, only few studies about calibration of 3-D probes were carried out. Poon and Rohling \cite{Poon05Comp3DCalib} compared 3-D calibration based on a IXI-fiducial wire phantom, a pointer tip phantom and a cube phantom. The IXI wire phantom and the cube phantom methods require only one volume acquisition for calibration. The presented feature detection is semi-automatic. The best results yielded the IXI phantom with a mean error in reproducibility of 1.5 mm, a RMS error of the point accuracy measure of 2.15 mm and a RMS error of the reconstruction accuracy by distance measure of 1.52 mm. Bouchet et al \cite{Bouchet01Calib} examined Z-fiducial phantom and achieved a RMS point accuracy error of 1.1mm. Two variants of a surface registration based 3-D calibration method were presented by Lange and Eulenstein in \cite{Lange02CalibSweptVolume3DUS}. The first one registers 3-D US images of the phantom with a geometric model derived from its CT scan. The second variant registers a number of US images of the phantom acquired from different positions. In both cases, surfaces are extracted manually. The authors claim that the latter approach could be fully automated. The CT variant performed best and yielded a RMS error in reproducibility precision of 1.8 mm and a RMS error in point accuracy of 2.0 mm. The ultrasound speed distortion problem is not addressed.

In this study we propose a 3-D US calibration method based on a single plane membrane phantom. A fast, precise and accurate 3-D feature extraction algorithm relying on the 2-D Hough transform is presented. In contrast to existing 3-D US calibration systems, the feature extraction process is fully automated. In the result section, precision and accuracy assessments are carried out using a specially designed validation phantom.

\section{Materials and Methods}
\subsection{Acquisition Hardware}
The acquisition hardware consists of a GE Voluson 730 Pro 3-D US scanner and a NDI Polaris optical tracker with a 0.25 mm RMS error (as communicated by NDI). The tracking system operates with wireless (passive) infrared-reflecting rigid bodies equipped with flat markers. The ultrasound volumes are acquired with a 5 to 9 Mhz two dimensional curved array probe (see Fig. \ref{fig:hardware}a). The piezo array of the probe is mounted on a mechanical device which is capable of sweeping regularly around its rotation axis within a predefined angular range. During the continuous sweeping process the US hardware reconstructs 3-D volumes from the series of acquired 2-D slices. The 3-D acquisition time ranges from 1s to 4s, depending mainly on sweep angle and axial acquisition depth. Images are digitally transferred using a proprietary software from GE Medical Systems named 4D View. The US scanner also communicates the voxel size. The scan converter assumes the speed of sound (SoS) in tissue to be 1540 m/s.

\subsection{The Membrane Phantom}
\begin{figure}
\centering
	\subfigure[]{
		\includegraphics[width=.354375\textwidth]{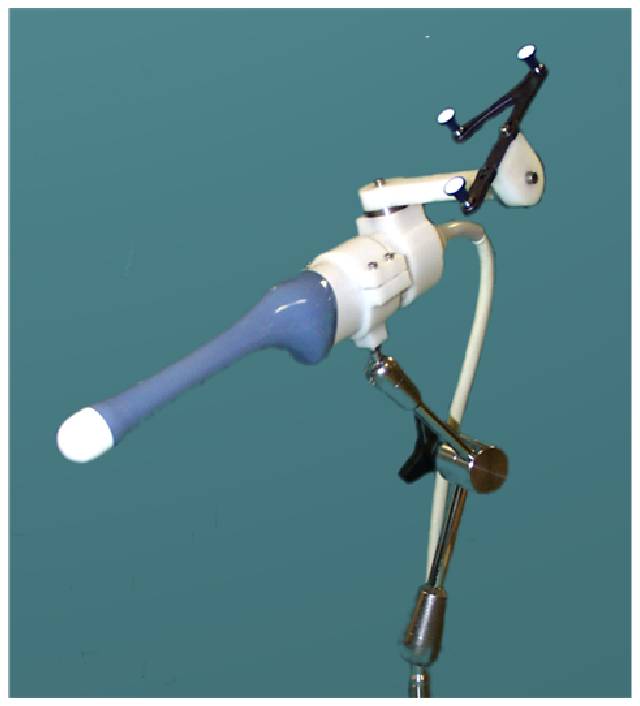}
	}
	\hfill
	\subfigure[]{
		\includegraphics[width=.545625\textwidth]{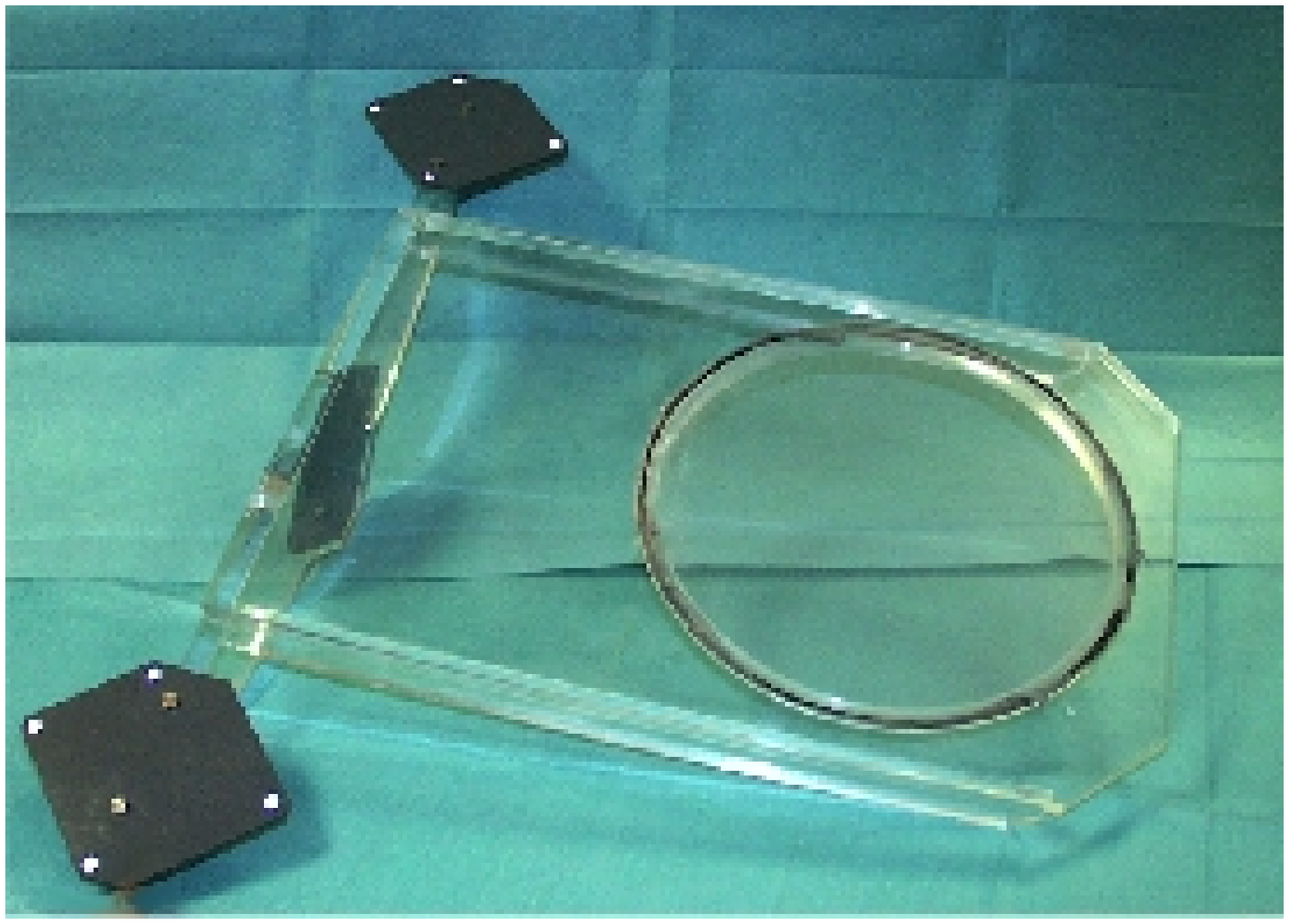}
	}
	\caption{Calibration hardware. Figure (a) shows an endorectal US probe mounted on an articulated arm. Figure (b) shows the membrane phantom. Both the probe and the membrane are equipped with infra-red reflecting passive rigid bodies for tracking.}
	\label{fig:hardware}
\end{figure}

The calibration phantom being dedicated to a clinical context, ergonomics considerations had an important impact on its design. We use a variation of the wall phantom presented in \cite{Prager98RapidCalibFreehand}, which is based on imaging the bottom wall of a water tank. The geometric form of the wall, which is a line in 2-D and a plane in 3-D, can be very robustly extracted from the US data using statistical algorithms like the Hough transform. This makes it possible to fully automate the feature extraction process without significant loss in precision and accuracy. This represents a big advantage over semi-automatic point-detection based phantoms in terms of calibration speed and ease of use. To overcome the plane thickness and the reverberation problems observable in US images of rigid surfaces \cite{Prager98RapidCalibFreehand} a filigrane nylon mesh membrane, tightly spanned on a planar rigid support with a circular and about 20cm wide hole, is used as target (see Fig. \ref{fig:hardware}b). Reverberation is further reduced by inclining the membrane plane with respect to the water tank bottom by 45°. A tracking reference (rigid body) is mounted on the membrane frame for phantom localization. The phantom is filled with water and equipped with a thermometer to measure water temperature.
\subsection{3-D Calibration Mathematics}
\begin{figure}
	\centering
		\includegraphics[width=.6\textwidth]{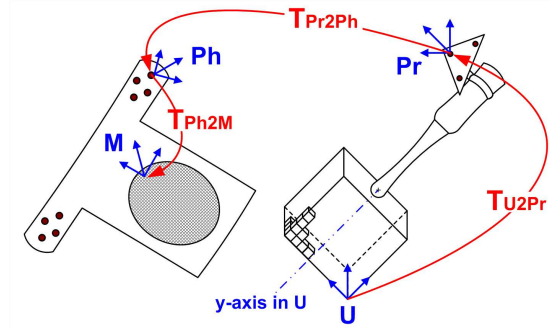}
	\caption{Illustration of the transformations involved in the calibration process. Note that scaling is omitted from the scheme for simplification.}
	\label{fig:Transformation_Scheme}
\end{figure}

As illustrated in Fig. \ref{fig:Transformation_Scheme}, four references are relevant for calibration: first of all, the membrane space $\mathbf{M}$ is defined as a reference in which the membrane lies in the origin and is parallel to $e_x$ and $e_y$ base vectors of $\mathbf{M}$. In this space, every point with a zero z-ordinate is a membrane plane point. The phantom space $\mathbf{Ph}$ and the probe space $\mathbf{Pr}$ are defined by the rigid bodies that are attached on the phantom and on the probe. Finally, the US volume space $\mathbf{U}$ corresponds to the voxel space of the 3-D images acquired by the ultrasound device. $\mathbf{T_{Ph2M}}$, $\mathbf{T_{Pr2Ph}}$ and $\mathbf{T_{U2Pr}}$ are homogenous 4x4 transformation matrices.

Suppose that we identified a point $\mathbf{p}=(x, y, z)$ in a US volume $\mathbf{U}$ as a point belonging to the membrane. With $\mathbf{s}=(s_x, s_y, s_z)$ denoting the voxel scale factors, it is verified that
\begin{equation}
\label{FundCalibEqn}
	\left(
		\begin{array}[1]{c}
			m_1 \\ m_2 \\ 0 \\ 1 
		\end{array}
	\right)
	=
	\mathbf{T_{Ph2M}} \cdot \mathbf{T_{Pr2Ph}} \cdot \mathbf{T_{U2Pr}} \cdot
	\left(
		\begin{array}[1]{c}
			s_x x \\ 	s_y y \\	s_z z \\	1
		\end{array}
	\right).
\end{equation}
where $\mathbf{T_{Ph2M}}$ is known from membrane pre-calibration (see chap. \ref{sec:precalib}) and  $\mathbf{T_{Pr2Ph}}$ is given by the tracking system. Further, the scaling vector $\mathbf{s}$ is communicated by the US hardware. The remaining unknown element is the homogenous rigid transformation $\mathbf{T_{U2P}}$. For convenience we define the elements of $\mathbf{T_{Pr2Ph}\cdot T_{Ph2M}}$ as $a_{ij}$ and the elements of $\mathbf{T_{U2P}}$ as $b_{ij}$ ($i,j \in 1..4$). The zero component of (\ref{FundCalibEqn}) yields then
\begin{equation}
\label{ZeroElemEqn}
\begin{array}{rcl}
0 & = & a_{31} ~( s_x x b_{11} + s_y y b_{12} + s_z z b_{13} + b_{14} )~+ \\
	&	  & a_{32} ~( s_x x b_{21} + s_y y b_{22} + s_z z b_{23} + b_{24} )~+ \\
	&	  & a_{33} ~( s_x x b_{31} + s_y y b_{32} + s_z z b_{33} + b_{34} )~+ \\
	&	  & a_{34}~.
\end{array}
\end{equation}
Using Euler angles and a three-dimensional vector we can represent $\mathbf{T_{U2P}}$ with six variables, which leaves us in total with 6 unknowns to solve for. A detected plane can be added to the equation system by adding at least three plane points (Using of course the $\mathbf{T_{Pr2Ph}}$ measured while acquiring the US volume in which the plane was detected).
%
\subsection{Membrane Pre-calibration}
\label{sec:precalib}
To reduce the number of degrees of freedom of the calibration process, the membrane space $\mathbf{M}$ is determined using a pointer equipped with a tracking reference. A large number of surface points of the membrane-supporting structure is acquired in order to compute the plane equation using a least square approximation combined with a simple M-estimator to increase robustness. Since the phantom rigid body is permanently fixed on the phantom, pre-calibration has to be carried out only once.
\subsection{Acquisition Protocol}
The major drawback of single-plane phantoms resides in their barely discriminative geometry. A plane can be described with only three variables, from which it follows that even with an optimal acquisition protocol, a minimum number of two acquisitions is necessary to cover all degrees of freedom. To obtain robust results the twelve-step acquisition protocol presented in \cite{Prager98RapidCalibFreehand} is used. The protocol improvement presented in \cite{Treece03HighDefinition} mainly addresses the z-axis imprecision problem inherent of most 2-D calibration systems. Since 3-D probes give as much information in z-direction as in x- or y- direction this modification yields no particular advantage in the 3-D domain but requires at least 18 steps. For that reason we stick to the original version. 

Sweeping and volume reconstruction being a continuous process of 1 to 4 seconds, significant distortions can be observed in the US volume when the probe is moved rapidly. Also, no direct access to the digital data is available which prevented synchronization of probe position measurement with US image acquisition. Therefore an articulated arm for complete probe immobilization during acquisition is used, eliminating all motion-induced artifacts and time lags. Furthermore, immobilizing the probe makes it possible to perform high precision position measuring based on a large number of measures and outlier elimination.
%
%
\subsection{Feature Extraction}
\begin{figure}
\centering
	\subfigure[]{
	\includegraphics[width=.35\textwidth]{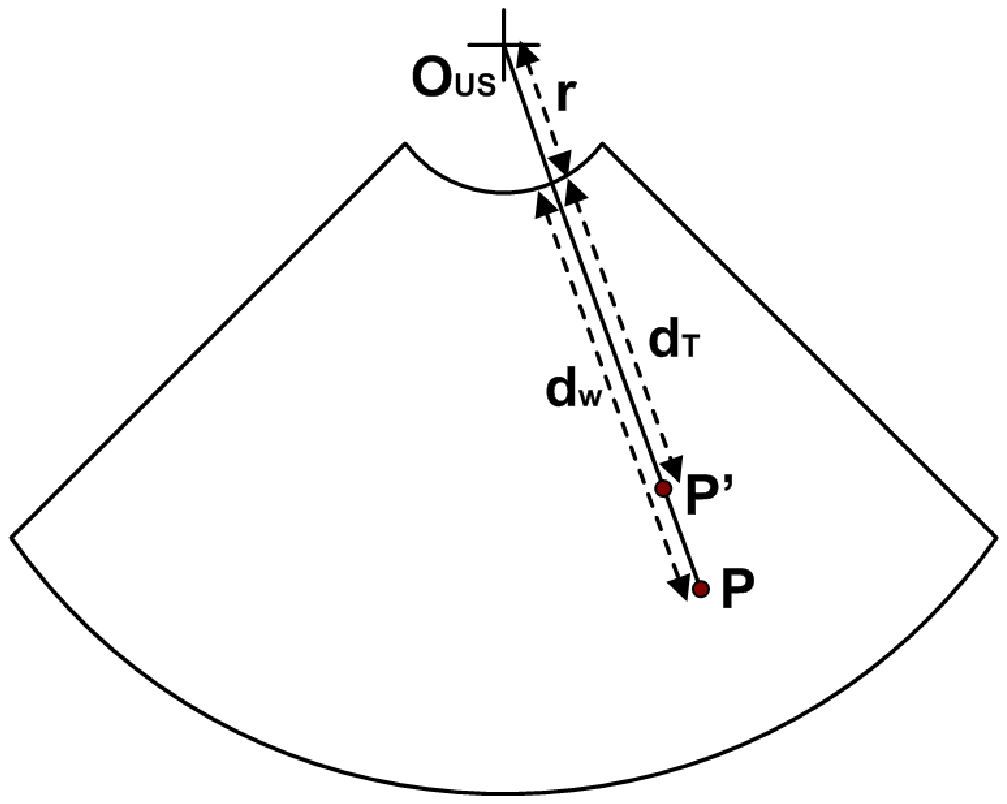}
	}
	\hspace{.1\textwidth}
	\subfigure[]{
	\includegraphics[width=.2625\textwidth]{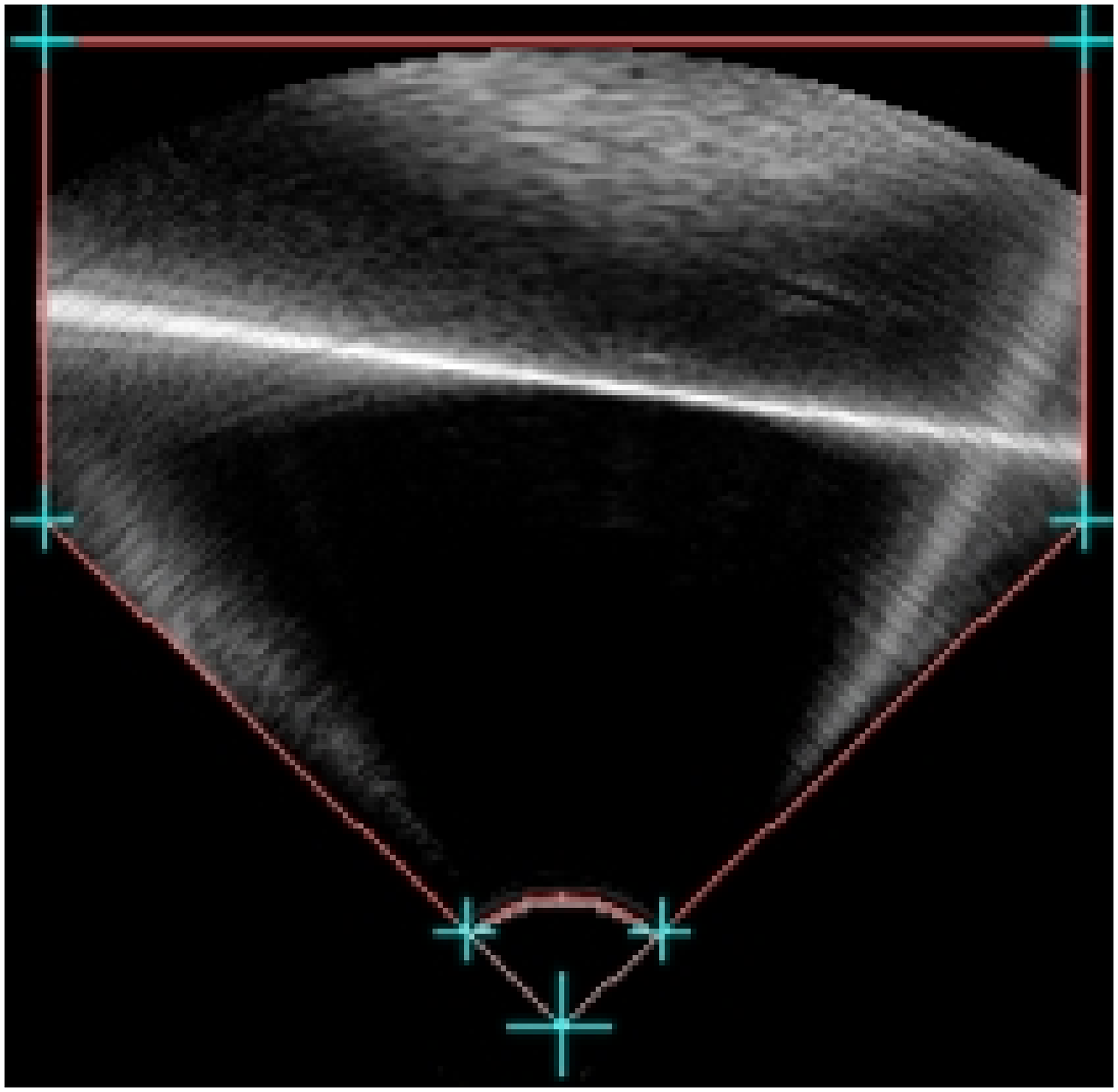}
	}
	\caption{US speed correction. Fig. (a) illustrates the correction geometry of a sectorial probe. $\mathbf{O_{US}}$ is the probe origin, $r$ the probe surface radius, $d_W$ the point's $\mathbf{P}$ distorted distance from the probe surface in water, $d_T$ the corrected distance and $\mathbf{P'}$ the corrected point. Fig. (b) shows the probe mask used to determine US origin and probe surface radius.}
	\label{fig:featextract}
\end{figure}

The first step of the feature extraction process consists in correcting the distortion caused by the difference between US speed in water at room temperature and in human tissue at 37°. To determine US speed in water in function of temperature the polynomial formula established by Bilaniuk and Wong was used \cite{Bilaniuk93Speed,Bilaniuk96ErratumSpeed}. A distortion geometry overview for all common probe types is given in \cite{Goldstein00Acoustic}. The distortion geometry of a sectorial probe is given in Fig. \ref{fig:featextract}a. With $v^t_W$ being the US velocity in water for a given temperature $t$ and $v_T$ being the velocity in tissue, $d_T$ is determined using the following formula:
\begin{equation}
d_T = \frac{v_T}{v_W}\cdot d_W.
\end{equation}
Sectorial probe speed correction requires manual definition of the US origin and the scan head surface radius. A graphical user interface was developed for this purpose (see Fig. \ref{fig:featextract}b). Origin and surface radius have to be defined only once during calibration. 
Plane detection can be carried out using the 3-D Hough transformation, but it would take several minutes to compute the result. Fortunately it is possible to determine the plane with good precision by simply extracting its intersection with two arbitrary volume slices, using the 2-D Hough transform. To facilitate and to accelerate US speed correction the $xy$ and $zy$ planes passing through the scan head origin were used. The Hough transform implementation uses intensity accumulation and the following threshold $s_H$ for an image $\mathbf{I}$:
\begin{equation}
s_H = \max\{i \in Hist(\mathbf{I})\} + (\max \{i \in \mathbf{I}\} - \min \{i \in \mathbf{I}\}) / 3~.
\end{equation}
The purpose of $s_H$ is to ignore the low-intensity water background, which represents the largest part of the image.
\begin{figure}
	\centering
		\subfigure[]{
		\includegraphics[width=.33\textwidth]{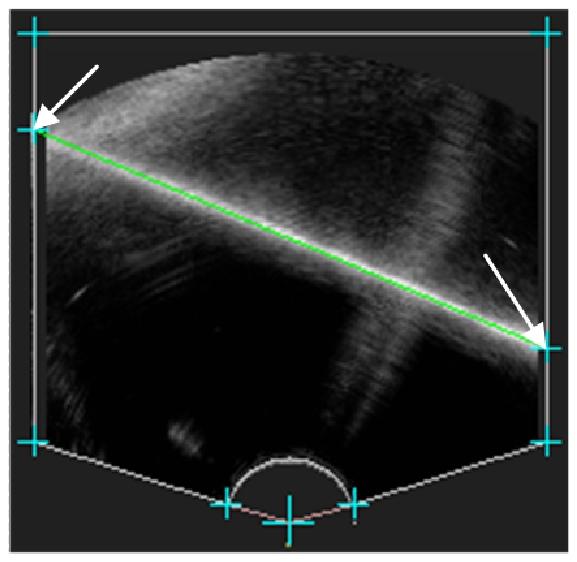}
		}
		\subfigure[]{
		\includegraphics[width=.33\textwidth]{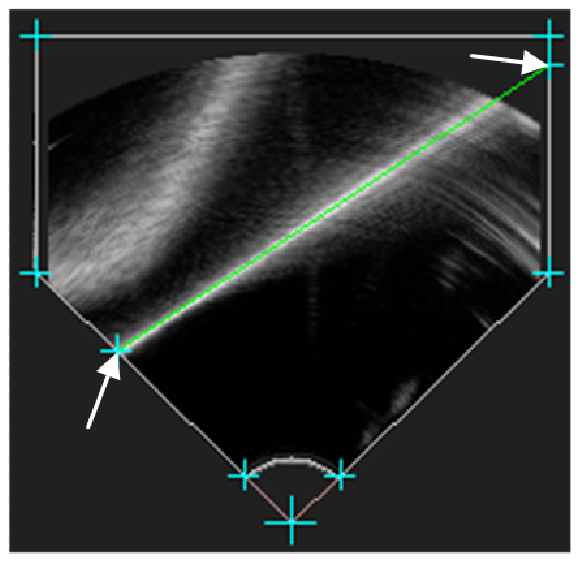}
 		}
	\caption{Screenshot of a successful automatic plane extraction. Note that lines are correctly detected in spite of a degraded membrane image caused by a steep scan angle. The arrows point at the line intersections with the mask.}
	\label{fig:Screenshot}
\end{figure}
\subsection{Optimization}
Optimization of (\ref{ZeroElemEqn}) is carried out with the non-linear Levenberg-Marquardt implementation given in \cite{NRecipies}. A random restart scheme within a range of reasonable initialization values robustifies this process. 
\subsection{Visual Back-tests}
The plane coordinates as resulting from the optimization process are visualized as a line in the slices used for feature extraction, and distance plus rotational errors between the segmented and the calculated line are evaluated. This allows for manual replacement of evident outliers with new acquisitions and for recomputation of the calibration without requiring a complete restart.
\section{Experimental Results}
\subsection{Test Configuration}
Precision and accuracy assessments were carried out using a membrane plane pre-calibration with an RMS surface distance error of 0.43mm for the measured surface points. A total of ten calibrations were performed using the twelve-step protocol. The probe rigid body was not moved between calibrations. The probe was mounted on an articulated arm and immobilized during position and image acquisition. The water temperature was 23°. The acquired US volumes had a size of (199, 199, 199) isotropic voxels with 0.477mm side lengths.

\subsection{Feature Extraction Quality}
The Hough transform extracted lines correctly for 238 out of 240 acquired images. Line detection failed for images on which only a very small part of the membrane was visible. In these cases, lines had to be manually determined. Note that no manual outlier elimination was carried out. As detection failures were rare, calibration was in average carried out in about twenty minutes. To get a better idea about the quality of feature extraction and about the presence of distortions in the membrane images we measured the detection precision: using (\ref{ZeroElemEqn}) we can calculate the distance error between a measured plane point and the computed plane as follows:
\begin{equation}
\label{ErrFormular}
\begin{array}{rcl}
\epsilon(x, y, z) & = & a_{31} ~( s_x x b_{11} + s_y y b_{12} + s_z z b_{13} + b_{14} )~+ \\
	&	  & a_{32} ~( s_x x b_{21} + s_y y b_{22} + s_z z b_{23} + b_{24} )~+ \\
	&	  & a_{33} ~( s_x x b_{31} + s_y y b_{32} + s_z z b_{33} + b_{34} )~+ \\
	&	  & a_{34}~.
\end{array}
\end{equation}
For each calibration, the average and the root mean square (RMS) distance of a set of points to the pre-calibration plane was computed using (\ref{ErrFormular}). For each line we computed ten equidistant points between the extreme points on the line segment inside the US volume. The angular feature extraction error is defined as the angle between the computed plane normal and the cross product of the directional vectors of the two extracted lines. Based on this definition the maximum and the RMS angular errors were computed for each acquired volume of the calibration. The aggregated errors for all calibrations can be found in Table \ref{IntraPrecision}.
\begin{table}
	\centering
	\caption{Aggregated Feature Extraction Precision.}
	\label{IntraPrecision}
	\begin{tabular}{lccc}
		\hline \vspace{-5pt} \\ 
		& \textbf{~~Distance Error~~} & \textbf{~~Distance Error~~} & \textbf{~~Angular Error~~} \\
		& [mm] & [vox] & [deg] \\ \hline \vspace{-5pt} \\
		\textbf{RMS Error~~} & 0.37 & 0.77 & 0.26 \\
		\textbf{Max Error~~} & 1.30 & 2.73 & 1.09 \\
	\end{tabular}
\end{table}
%
%
\subsection{Calibration Precision}
The calibration precision measures the reproducibility of calibration results. Again, both the translational and the angular errors were assessed. The translational error is defined as the standard deviation of the volume center after scaling and right-hand multiplication to the different calibration transformations $\mathbf{T^i_{U2Pr}}$. The angular error is measured as the standard deviation of angular differences between the $(0,0,1)$ vector after scaling and right-hand multiplication to the different calibration transformations $\mathbf{T^i_{U2Pr}}$ (see Table \ref{InterPrecision}).
%
%
%
%
%
%
%
%
%
\begin{table}
	\caption{Calibration Precision.}
	\centering
	\label{InterPrecision}
	\begin{tabular}{lccc}
		\hline \vspace{-5pt}\\
		& \textbf{~~Distance Error~~} & \textbf{~~Distance Error~~} & \textbf{~~Angular Error~~} \\
		& [mm] & [vox] & [deg] \\ \hline \vspace{-5pt} \\
		\textbf{Standard Deviation~~} & 1.15 & 2.41 & 0.61 \\
		\textbf{Max Error~~} & 1.99 & 4.03 & 1.12 \\
	\end{tabular}
\end{table}

\subsection{Reconstruction Accuracy}

\begin{figure}
	\centering
		\includegraphics[width=.5\textwidth]{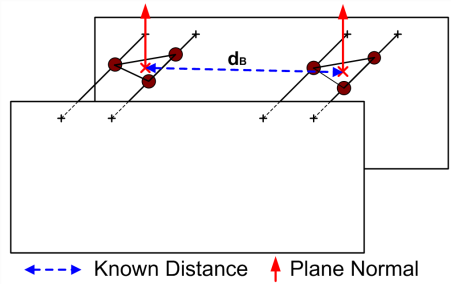}
	\caption{Reconstruction Accuracy Measurement Phantom.}
	\label{fig:AccuracyPhantom}
\end{figure}

Reconstruction accuracy was assessed using the bead phantom illustrated in Fig. \ref{fig:AccuracyPhantom}. Note that the beads are co-planar within a precision of 0.25mm (RMS). The left-hand three beads form the left triangle while the right-hand beads form the right triangle. The distance $d_B$ of the triangle barycenters was evaluated with an estimated accuracy of about 0.5mm. 

Twenty US images of the phantom were acquired, ten imaging the left triangle and ten the right triangle. Note that images of one triangle did not intersect with the second one. The bead centers were manually extracted from the images. The reconstructed triangle barycenters and normals were then projected into probe space for each calibration, which yielded 100 point and vector pairs. The distance error for a point pair is defined as the difference between their Euclidean distance and $d_B$. The angular error for a vector pair is defined as the angle between both vectors. The results are given in Table \ref{AccuracyErr}.

\begin{table}
	\caption{Reconstruction Accuracy Results.}
	\label{AccuracyErr}
	\centering
	\begin{tabular}{lccc}
	\hline \vspace{-5pt}\\
		 & \textbf{~~Distance Error~~} & \textbf{~~Distance Error~~} & \textbf{~~Angular Error~~} \\ 
		 & [mm] & [vox] & [deg] \\ \hline \vspace{-5pt} \\
		\textbf{RMS Error~~} & 0.90 & 1.90 & 1.79 \\
		\textbf{Max Error~~} & 2.44 & 5.11 & 3.03 \\ \hline
	\end{tabular}
\end{table}

\section{Discussion}

Probe calibration for currently available swept-volume 3-D probes makes only sense for static applications. Real-time access to acquired 3D images is currently not provided by the Voluson hardware. Also, depending on scan parameters, the duration of the volume sweeping process ranges from 1s to 4s. Due to probe or tissue motion, the physical location of voxels can therefore be way off the position indicated by the calibration. The latter problem could be reduced by scan-heads equipped with high-frequency sweeping-devices, but it will disappear only with non-sweeping 2-D piezo array US probes. Until then, it is better to calibrate probes using an articulated arm for immobilization.

Passing from 2-D to 3-D calibration improves calibration results (for information on 2-D precision of the membrane phantom see \cite{Hook03Calibration}) because the z-axis-uncertainty inherent of 2-D calibration is eliminated: when giving a plane instead of a line as input to the optimizer the rotational degrees of freedom are significantly better covered. This allowed us to reduce the number of acquisitions for calibration while still achieving precise and accurate results.

Feature extraction from membrane phantom images showed both robust and precise results. On our set of test images the line blurring and intensity degradation effects occurring when scanning a wall phantom from an oblique angle were correctly handled by the Hough transform: lines were consistently placed in the center of the beam width. Note that the feature extraction precision RMS and maximum errors reported in Table \ref{IntraPrecision} are relatively small, which indicates that the physical plane location corresponds indeed with the beam width center line. Membrane reverberation was not observable and did therefore not disturb the detection process. Also, the membrane phantom was not exposed to line thickness problems. Due to these characteristics, feature extraction could be fully automated (up to the manual US origin and probe radius determination required for US speed correction).

The user-independency resulting from automated feature extraction is partially coun\-ter-balanced by the necessity to follow a protocol for data acquisition, which can re-introduce user bias. Nevertheless we believe that it is more convenient to follow a simple acquisition protocol instead of extracting features semi-automatically from US volumes. Further we preferred to correct for US speed errors instead of requiring 50° water or phantom fill materials that have the same US characteristics than human tissue, which makes it possible to use the phantom in a clinical context.

The overall calibration time of about 20 minutes is mostly due to manipulation of the articulated arm, high precision probe position measurement and data transfer, which requires several manual interventions. Feature extraction and optimization is computed in several seconds. In cases where the feature extraction precision evaluation shows poor results, a visual verification and eventually a correction have to be carried out, which requires some additional minutes. Significant speed-up could be achieved by automating the communication between the US scanner and the calibration computer.

The presented calibration system assumes that the SoS in human tissue is uniform and that it corresponds to the SoS internally used by the US scanner, which is in general the mean SoS in tissue of 1540m/s. However, SoS varies with different types of tissue: The SoS in fat is approximately 1450m/s, in blood 1570m/s, in the brain 1541m/s and in water 1480m/s. As the US generally crosses tissue layers of different thickness and different types on its way through the body, and as the target tissue is often viewed from various positions, the in vivo accuracy of the calibration may show fluctuations of more than 5 per cents in extreme cases during an examination. Also, for some applications it would be appropriate to use a different mean SoS than the 1540m/s for calibration, but this is beyond the scope of this study.

Future work will address the twelve-step acquisition protocol which contains a lot of redundancy. Also, our system does currently not provide a foolproof indicator for missing coverage of degrees of freedom. We therefore started experiments with an Eigenvalue system similar to the one presented in \cite{Hsu05RapidFreehandCalib}.
\section{Conclusion}
A robust 3-D US probe calibration method designed for clinical usage was presented. Calibration can be carried out in about twenty minutes due to fully automatic 3-D Plane extraction based on robust and efficient 2-D line detection. The point reconstruction accuracy of our phantom can compete with previously presented 3-D phantoms: Lange and Eulenstein communicated RMS errors between 2.0mm and 2.2mm \cite{Lange02CalibSweptVolume3DUS}, Bouchet et al were confronted to 1.1mm RMS point accuracy \cite{Bouchet01Calib} while Pohn and Rohling published errors between 1.52mm for their IXI-wire, 1.59mm for the cube and 1.85mm for their stylus approach. With 0.90mm RMS point accuracy (see Table \ref{AccuracyErr}) we achieved slightly better results. Finally, the proposed method is temperature-independent and uses water as transmission matter which facilitates its usage.
\newline\newline\noindent\small{\textbf{Acknowledgements:} This project is supported by PRAXIM/KOELIS, ANRT and PHRC 2003 "Prostate-echo" (from French Health Ministry) grants.}

\bibliographystyle{splncs}
\bibliography{Calibration}

\end{document}